\documentclass[aps,prl,twocolumn,preprintnumbers,superscriptaddress,10pt]{revtex4-1}
\usepackage[mathscr]{eucal}
\usepackage{epsf,color,amsmath,amssymb,amsfonts}
\usepackage{enumerate}
\usepackage{hyperref}
\usepackage{graphicx}
\usepackage{psfrag}
\usepackage{float}
\usepackage{dcolumn,bm}

\newcommand{\bea}{\begin{eqnarray}}
\newcommand{\beal}[1]{\begin{eqnarray}\label{#1}}
\newcommand{\eea}{\end{eqnarray}}
\newcommand{\be}{\begin{equation}}
\newcommand{\bel}[1]{\begin{equation}\label{#1}}
\newcommand{\ee}{\end{equation}}

\newcommand{\bit}{\begin{itemize}}
\newcommand{\eit}{\end{itemize}}
\newcommand{\ben}{\begin{enumerate}}
\newcommand{\een}{\end{enumerate}}

\newcommand{\mt}[1]{\textrm{\tiny #1}}
\newcommand{\Thyd}{T_\mt{hyd}}
\newcommand{\thyd}{t_\mt{hyd}}
\newcommand{\nc}{N_\mt{c}}

\newcommand{\eqn}[1]{(\ref{#1})}
\newcommand{\ellc}{\ell_\mt{char}}

\newcommand{\Fig}[1]{Fig.~\ref{#1}}

\begin{document}
\onecolumngrid
\preprint{ICCUB-13-245}

\title{Longitudinal Coherence  in a Holographic Model of Asymmetric  Collisions}
\author{Jorge Casalderrey-Solana} 
\affiliation{Departament d'Estructura i Constituents
de la Mat\`eria, Institut de Ci\`encies del Cosmos,
Universitat de Barcelona, Mart\'\i \ i Franqu\`es 1, 08028 Barcelona, Spain}
\author{Michal P.~Heller}
\altaffiliation[On leave from: ]{\emph{National Centre for Nuclear Research,  Ho{\.z}a 69, 00-681 Warsaw, Poland.}}
\affiliation{Instituut voor Theoretische Fysica, Universiteit van Amsterdam \\
Science Park 904, 1090 GL Amsterdam, The Netherlands}
\author{David Mateos}
\affiliation{{Instituci\'o Catalana de Recerca i Estudis Avan\c cats (ICREA), 
Barcelona, Spain}}
\affiliation{{Departament de F\'\i sica Fonamental,  Institut de Ci\`encies del Cosmos, Universitat de Barcelona, Mart\'{\i}  i Franqu\`es 1, E-08028 Barcelona, Spain}}
\author{Wilke van der Schee}
\affiliation{Institute for Theoretical Physics and Institute for Subatomic Physics,
Utrecht University, Leuvenlaan 4, 3584 CE Utrecht, The Netherlands}


\begin{abstract}
\noindent
As a model of the longitudinal structure in heavy ion collisions, we simulate gravitational shock wave collisions in anti-de Sitter space in which each shock is composed of multiple constituents. We find that all constituents act coherently, and their separation leaves no imprint on the resulting plasma, when this separation is $\lesssim 0.26 / \Thyd$, with $\Thyd$ the temperature of the plasma at the time when hydrodynamics first becomes applicable. In particular, the center-of-mass of the plasma coincides with the center-of-mass of \emph{all} the constituents participating in the collision, as opposed to the center-of-mass of the individual collisions. We discuss the implications for nucleus-nucleus and proton-nucleus collisions.

\end{abstract}
\maketitle

\noindent
{{\bf 1. Introduction.}}
The hydrodynamic behaviour of the matter produced in the high energy collision of two large nuclei  is one of the most striking  results of the heavy ion programs at RHIC and LHC \cite{Adcox:2004mh, Adams:2005dq, Muller:2012zq}. One of the theoretical challenges posed by this collective behavior is to understand the hydrodynamization process: the transition from the initial far-from-equilibrium regime to the regime that is well described by hydrodynamics. Interesting insights have been obtained through the dual gravitational description, in which a central nucleus-nucleus (A+A) collision has been toy-modeled as a symmetric collision 
in anti-de Sitter space (AdS) of two gravitational shock waves of infinite extent in the transverse directions  \cite{Chesler:2010bi,Casalderrey-Solana:2013aba,Chesler:2013lia, vanderSchee:2014qwa}. 
For an extensive review of applications of the gauge/gravity duality to heavy ion physics, see \cite{CasalderreySolana:2011us}.

Strong collective behavior may also occur in high-multiplicity proton-nucleus (p+A) and deuteron-nucleus (d+A) collisions. The recent analyses of p-Pb data from LHC \cite{Aad:2013fja,Chatrchyan:2013nka,Abelev:2012ola} and d-Au data  from RHIC \cite{Adare:2013piz,Sickles:2013yna} have shown flow signals in high-multiplicity events. While their interpretation as a hydrodynamic response is still far from settled, early hydrodynamic simulations seem 
to reproduce most of the observed systematics \cite{Bozek:2013uha}. This possibility motivates us to consider a holographic set-up that captures one of the key features of a (p/d)+A collision: the asymmetry in the longitudinal extents of the two projectiles \cite{footnote1}. Incorporating the different (and finite) extents in the transverse directions is certainly important but  technically harder and we leave it for future work. 

\begin{figure*}
\begin{tabular}{cc}
\includegraphics[width=0.49 \textwidth]{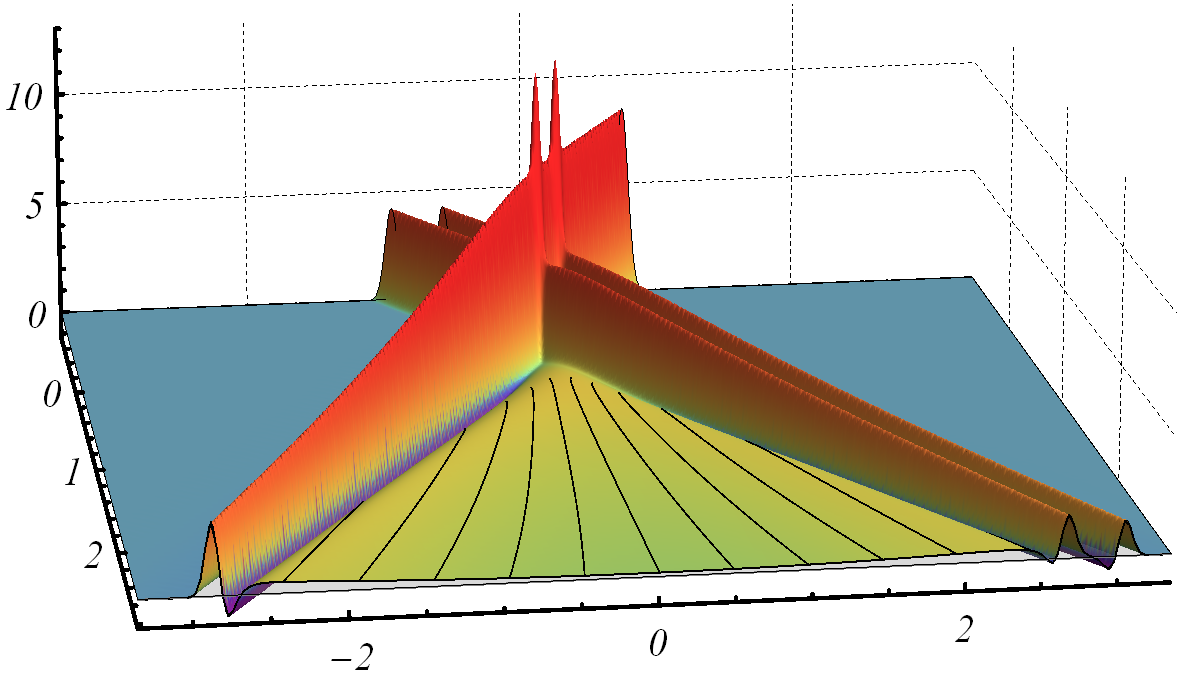}
\quad & \quad
\includegraphics[width=0.49 \textwidth]{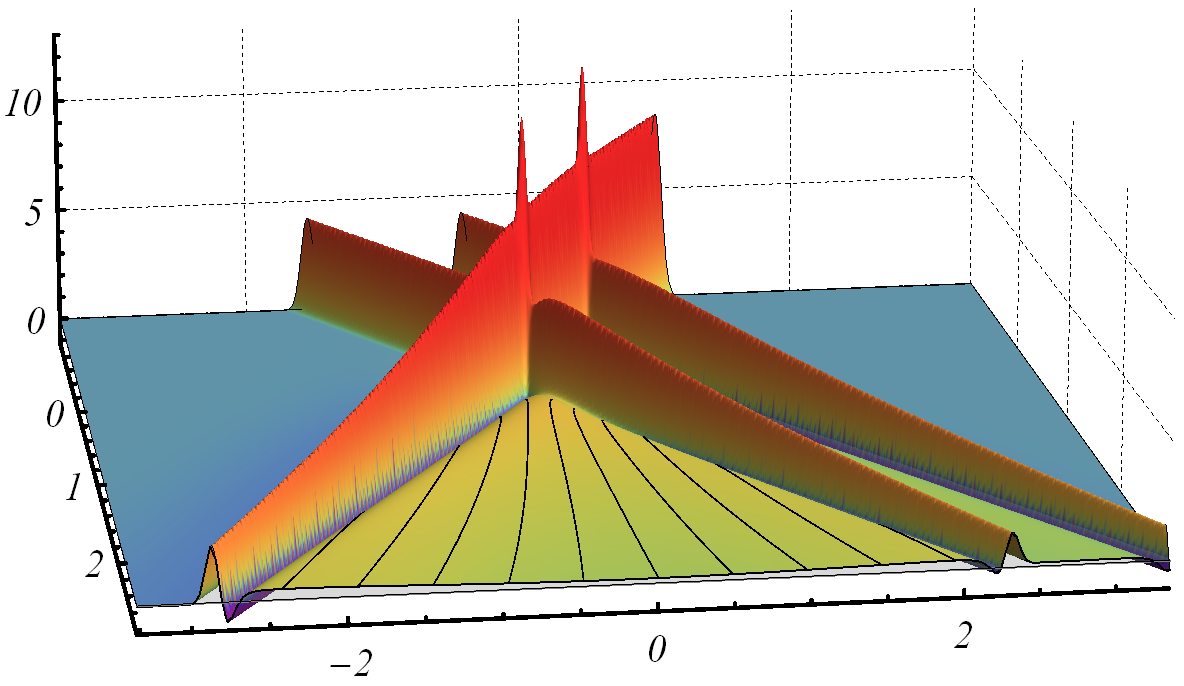}
\end{tabular}
\vspace{-0.1cm}
\caption{Energy density (divided by $N_c^2 / 2\pi^2$) of two asymmetric collisions. The black lines are streamlines of the produced plasma. 
}
\label{EnergyDensity}
\end{figure*}

We collide planar gravitational shocks in AdS$_5$ with different longitudinal profiles, which via the gauge/gravity duality provides a model for a high-energy collision of projectiles with non-trivial longitudinal structure \cite{footnote2}. In the  field theory, the shocks are dual  to two infinite sheets of energy characterised by a stress tensor whose only non-zero components are
\mbox{$T_{\pm\pm} (z_\pm)= 
(N_c^2 / 2\pi^2) F_\pm(z_\pm)$}, where $z_\pm=t \pm z$, $z$ is the beam direction, and $F_\pm(z_\pm)$ are two arbitrary profile functions associated with the left- and right-moving shocks, respectively. 
We choose $t=z=0$ as the point at which the center of masses (c.o.m.) of both shocks coincide.
The general expression for the shocks that we consider is of the form
\be
\label{eq:shock}
F_\pm=\frac{\mu^3}{w \sqrt{8\pi}} 
\left\{ \exp \left[ \frac{\left(z_\pm - \frac{1}{2}\ell_{\pm}\right)^2}{2w^2}\right] + 
\exp \left[ \frac{\left(z_\pm + \frac{1}{2}\ell_{\pm}\right)^2}{2w^2}\right]
\right\}\,.
\ee
This describes a `double shock' (with two Gaussian constituents) of characteristic size $\ellc \simeq \ell_{\pm}$ if 
$\ell_{\pm} \gg w$, and a `single shock' of characteristic size $\ellc \simeq 3.3 w$ (the region where 90\% of the energy is contained) if $\ell_{\pm}=0$. 
Each constituent is meant to be a cartoon of a nucleon participating in the collision. By varying $\ell_+$ and $\ell_-$ we can therefore model symmetric collisions (single-single and double-double collisions) and asymmetric collisions (single-double collisions). Note that we work in the c.o.m.~frame of the collision, in which each shock has the same energy per unit transverse area, $(N_c^2 / 2\pi^2) \mu^3$, regardless of the number of constituents. Because of conformal invariance, each of the shocks is characterised by the dimensionless products $\mu w$ and  $\mu \ell_\pm$.

Our main result is that the created plasma at mid-rapidity is insensitive to the structure of the initial shocks if the characteristic size of each shock satisfies \mbox{$\ellc \lesssim 0.26/\Thyd$}, where $\Thyd$ is the plasma temperature at the time of hydrodynamization, $\thyd$. We will refer to this feature as `longitudinal coherence'. 
In particular, even though the initial projectiles may be very asymmetric, in the coherent regime the c.o.m.~of the created plasma coincides with the c.o.m.~of \emph{all} the nucleons participating in the collision, as opposed to the c.o.m.~of each individual nucleon-nucleon collision.

\noindent
{{\bf 2. Longitudinal coherence.}} 
\Fig{EnergyDensity} shows the energy density for the two asymmetric collisions in the second row of Table \ref{TableofShocks}: a coherent  collision with $\ellc \simeq 0.12/\Thyd$ (left) and an incoherent collision with $\ellc \simeq 0.36/\Thyd$ (right). All constituents have $\mu w=0.05$, i.e.~they are `thin' in the language of \cite{Casalderrey-Solana:2013aba}. In both cases  the left-moving shock is a single-shock, while the right-moving shock is a  double-shock with $\mu \ell_-=8 \mu w=0.4$ (left) and $\mu \ell_-=24 \mu w=1.2$ (right). As expected from \cite{Casalderrey-Solana:2013aba}, the thin constituents pass through each other virtually undisturbed and then start to attenuate.
Close to the light-cone,  both figures show one left-moving and two right-moving attenuating maxima after the collision, indicating that in both cases the high-rapidity region is sensitive to the initial structure of the shocks.

\begin{figure*}
\begin{tabular}{cc}
\includegraphics[width=0.49 \textwidth]{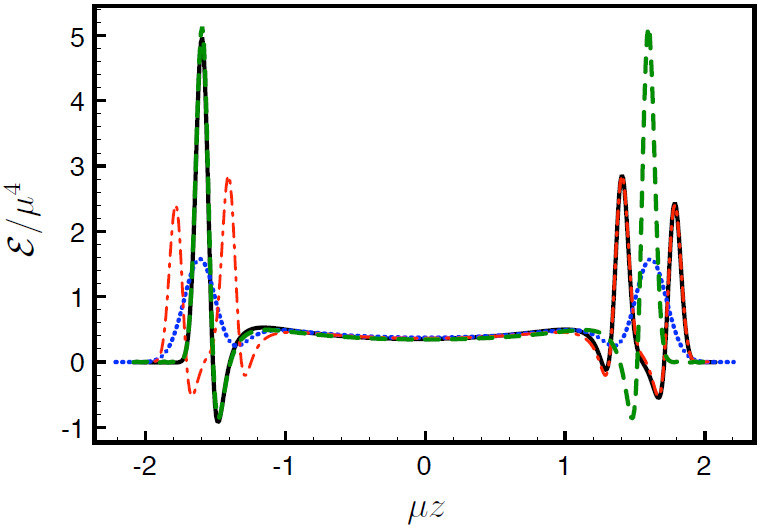}
\quad 
&
 \quad
\includegraphics[width=0.49 \textwidth]{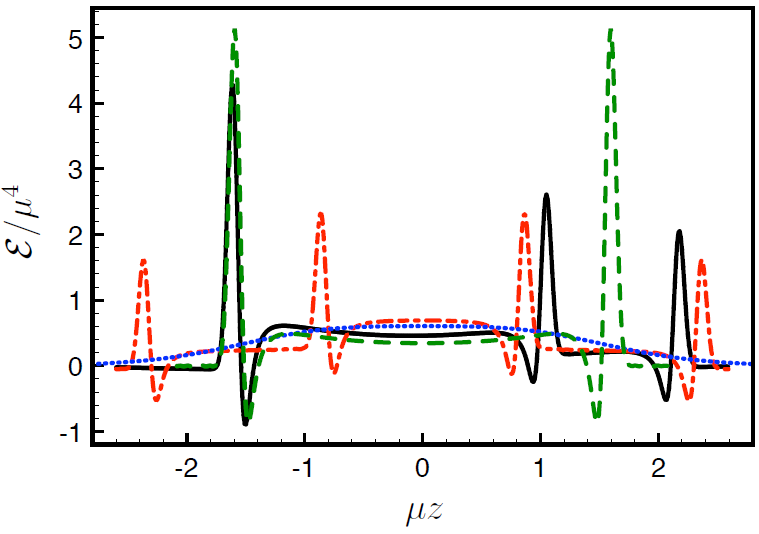}
\end{tabular}
\vspace{-0.2cm}
\caption{Energy density (divided by $N_c^2 / 2\pi^2$) at $\mu t=1.6$ for different shock collisions characterised by the parameters  displayed in Table \ref{TableofShocks}. 
}
\label{ShockComparison}
\end{figure*}

\begin{table*}[!ht]
\vspace{-0.2cm}
\caption{\label{TableofShocks} Parameters of the shocks displayed in \Fig{ShockComparison}.}
\begin{center}
\begin{tabular}{|l||c|c|c|c|c|c||c|c|c|c|c|c|c|c|c|c|c|c|} 
\hline
&
 \multicolumn{6}{c||} {Left} &  \multicolumn{6}{c|} {Right}
\\
\hline
 & $\mu  w$ & $\mu \ell_+$ &$\mu \ell_-$&$\mu \thyd$ &$\Thyd/\mu$ & 
 $\ellc \Thyd$ & $\mu  w$ & $\mu \ell_+$ &$\mu \ell_-$&$\mu \thyd$&$\Thyd/\mu$ & $\ellc \Thyd$ \\
\hline
Green-dashed (single-single) &  0.05 & 0   &   0   & 0.88   & 0.30  & 0.05 &  0.05  &   0 &  0 & 0.88 & 0.30 & 0.05\\
\hline
Black-continuous  (single-double) &  0.05 & 0   &   0.4 & 0.88 & 0.30 & 0.12 & 0.05 & 0 &         1.2 & 0.95 & 0.31 & 0.36 \\
\hline
Red-dotted-dashed (double-double) & 0.05 & 0.4 & 0.4 & 0.88 & 0.30 & 0.12 &  0.05 & 1.6 & 1.6 & 1.20 & 0.33 & 0.48 \\
\hline
Blue-dotted (single-single) & 0.10 & 0 & 0 & 0.88   & 0.30 & 0.1 &1.9 & 0 & 0 & -0.08 & 0.30 & 1.9  \\
\hline
\end{tabular}
\end{center}
\label{table}
\vspace{-0.5cm}
\end{table*}

In contrast, the mid-rapidity region of \Fig{EnergyDensity}(left) keeps no memory of the initial structure of the shocks. This is illustrated in \Fig{ShockComparison}(left), which shows snapshots of the energy density at a fixed time after hydrodynamization, $\mu t=1.6$, for the several collisions with different initial shock structures but with the same total energy listed in the left part of Table \ref{TableofShocks}. We see that the energy density around mid-rapidity for the single-double collision of \Fig{EnergyDensity}(left) is identical to that for a single-single or a double-double collision with constituents of the same width, and for a single-single collision with twice-as-thick constituents. In all these cases the hydrodynamization time and the hydrodynamization temperature are independent of the initial structure of the shocks. For single shocks this  is consistent with \cite{Casalderrey-Solana:2013aba}, where it was found that the hydrodynamization properties of the plasma are independent of the widths of the initial shocks provided these satisfy $\mu w \lesssim 0.2$. 

\Fig{ShockComparison}(right) shows  analogous snapshots for the collisions listed on the right part of Table \ref{TableofShocks}, which again have the same total energy but differ in the initial structure of the shocks.  One of the curves is the same single-single collision of thin shocks from \Fig{ShockComparison}(left), which is included for comparison. The other three curves all have  $\ellc > 0.26 / \Thyd$ and they illustrate the incoherent regime, namely the fact that the energy density around mid-rapidity, as well as the hydrodynamization time and the hydrodynamization temperature, are sensitive to the initial structure of the shocks. Note that the different hydrodynamization temperatures would translate into about a 30\% difference in the energy density at mid-rapidity (which scales roughly as $\Thyd^4$) even if each of these curves were plotted at its corresponding hydrodynamization time. 

From the gauge theory viewpoint, these results imply that the smallest longitudinal structure that the fields in the mid-rapidity region can resolve is set by the inverse temperature at hydrodynamization, which in the coherent regime is $\Thyd = 0.3\mu$. Clearly, the plasma will be sensitive to the structure of the initial shocks if their characteristic size, $\ell_\mt{char}$, is larger than the formation time of the hydrodynamized plasma, $\thyd$. By inspection of Table \ref{TableofShocks} we see that the transition between the coherent and the incoherent regimes takes place at a  scale $\ell_\mt{coh}$ such that $0.12 < \ell_\mt{coh} \Thyd < 0.36$. Since this transition is smooth, $\ell_\mt{coh}$ is not sharply defined. Motivated by the considerations above, we therefore choose to define it as the hydrodynamization time for single-single collisions of thin shocks, which yields $\ell_\mt{coh}=0.26/\Thyd$.

\begin{figure*}[t!]
\begin{tabular}{cc}
\includegraphics[width=0.47 \textwidth]{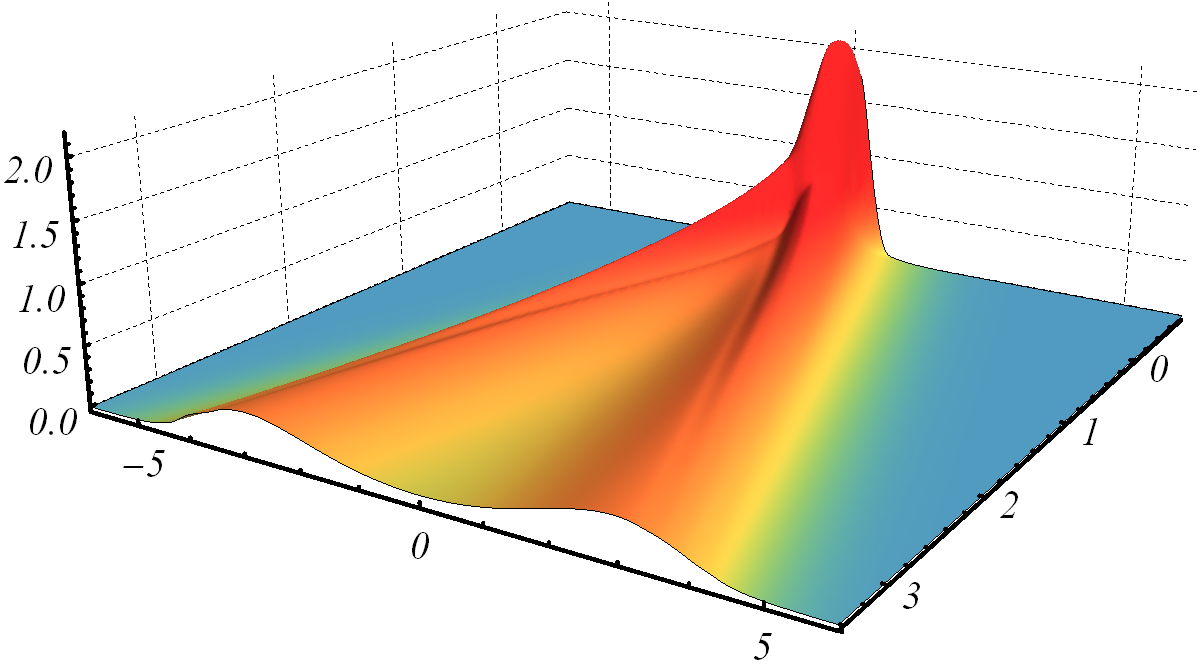}
\quad & \quad
\includegraphics[width=0.47 \textwidth]{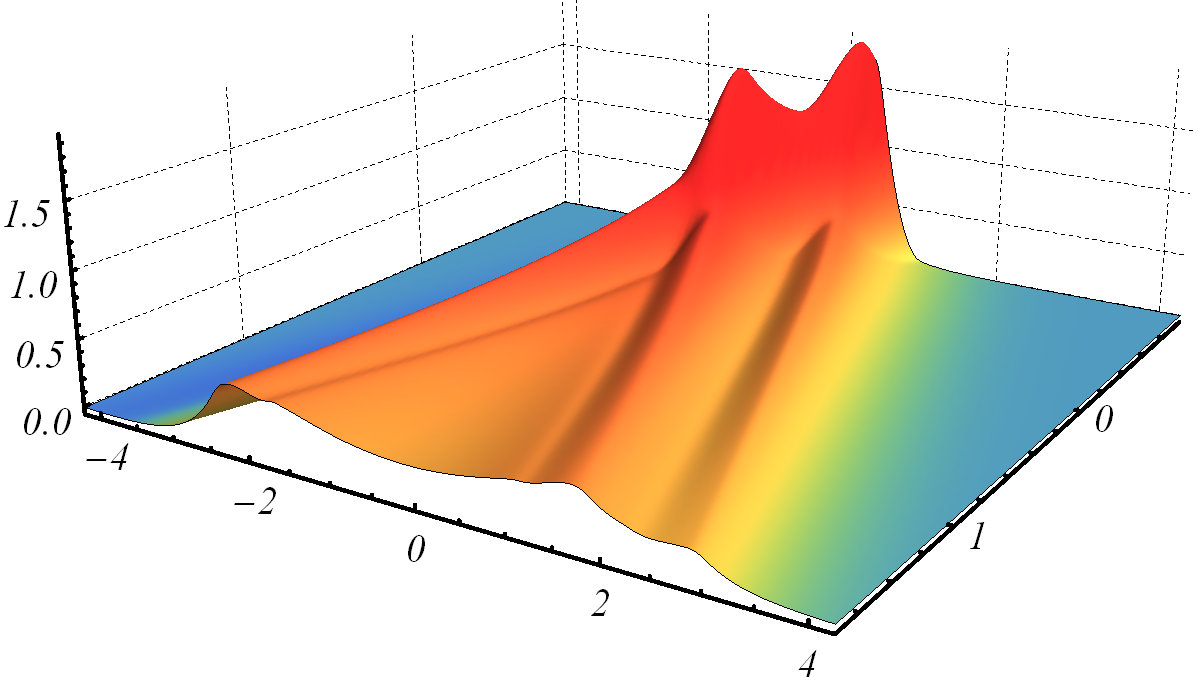}
\end{tabular}
\caption{
Volume element $\Sigma^3$ on the apparent horizons formed in the two collisions depicted in \Fig{EnergyDensity}, with $L$ the asymptotic AdS radius. In equilibrium the quantity plotted would be related to the gauge theory entropy density $s$ through 
\mbox{$\Sigma^3/\mu^3 L^3 = 2\pi s/\nc^2 \mu^3$}. The apparent horizon is computed as in \cite{Chesler:2010bi,Casalderrey-Solana:2013aba,Chesler:2013lia}.
}
\label{horizon}
\end{figure*}

This picture is supported by the gravitational description. In \Fig{horizon} we show the volume element on the apparent horizon formed in the two collisions displayed in \Fig{EnergyDensity}. Although this quantity depends on the slicing of the space-time, close to equilibrium it provides a lower bound for the entropy density \cite{Figueras:2009iu}. According to the gauge/gravity duality, the horizon encodes the physics at the thermal scale. Heuristically, one may say that \Fig{horizon} provides an effective picture of  \Fig{EnergyDensity} in which all length scales shorter than the thermal scale have been integrated out. 
It is therefore suggestive that in \Fig{horizon}(left) there is no trace of the microscopic structure of the shocks even at the time $t=0$ of the collision. In contrast, for the further-separated colliding shock constituents of  \Fig{EnergyDensity}(right), the corresponding apparent horizon in \Fig{horizon}(right) reflects the initial configuration, albeit with a significant smoothening due to the integration of scales.

\noindent
{{\bf 3. Discussion.}} 
Since longitudinal coherence only depends on the inability of the horizon to resolve sub-thermal length scales, we expect this coherence to occur in holographic high-energy collisions more general than the simple model considered here. These may include collisions of shocks with profiles more general than \eqn{eq:shock} and collisions with non-trivial transverse dynamics, at least if the transverse expansion rate is slower than the longitudinal one. In the following we take this as an assumption and explore 
an interesting consequence for high-multiplicity (p/d)+A collisions. Furthermore, we consider the limits in which the physics of bulk-particle production is assumed to be exclusively strongly or weakly coupled, 
the hope being that these limits bracket the production dynamics at the energies of present colliders. 

In the strong-coupling limit our results, together with the large Lorentz contraction of the colliding projectiles at RHIC and LHC, suggest that most of the participating nucleons act coherently in the formation of the plasma. As a consequence, the momentum rapidity of the plasma's c.o.m.,  $y_\mt{plasma}$, should coincide with the momentum rapidity of the c.o.m.~of all the participating nucleons, $y_\mt{part}$. 
Since the local energy density at fixed proper time is maximal at $y_\mt{plasma}$ \cite{Casalderrey-Solana:2013aba,Chesler:2013lia}, the maximum in the rapidity distribution of particles, $y_\mt{max}$, also coincides with  $y_\mt{part}$. For a generic collision with $N_\mt{A}$ ($N_\mt{B}$) right-moving (left-moving) participating nucleons moving at rapidity $y_\mt{A}$ ($y_\mt{B}$), we have that 
\mbox{$y_\mt{part}= \tfrac{1}{2}\,\log(N_\mt{A}/N_\mt{B}) +y_\mt{NN}$}, where $y_\mt{NN}=\tfrac{1}{2}(y_\mt{A}+y_\mt{B})$ is the rapidity of the nucleon-nucleon c.o.m.

This shift has interesting consequences for A+A collisions. Firstly, event-by-event fluctuations in the number of participating nucleons in A+A collisions lead to  fluctuations in $y_\mt{max}$ according to $y_\mt{part}$, as was also studied in \cite{Bzdak:2012tp}. Secondly, for off-central collisions there will be a similar shift locally in the transverse plane. Lastly, although in this Letter we focused on the plasma formed at mid-rapidity, it would also be interesting to study in more detail the high-rapidity region, where universal scaling (`limited fragmentation') is observed in both A+A \cite{Back:2002wb} and p(d)+A collisions \cite{Back:2004mr}. However, since addressing this feature would require a more sophisticated model that incorporates confinement and finite-$\nc$ effects, we leave this study for future work.

Perhaps even clearer is an observable consequence for p(d)-A collisions. There $y_\mt{max}$ shifts to the A side due to the asymmetric collision geometry. Taking $N_{A}=15-30$ as representative values for central p(d)+A collisions at the LHC (RHIC) we find $y_\mt{max}=0.9 \,(1.3) - 1.2\, (1.7)$. An additional result of the strong-coupling model is that the plasma is $y$-reflection-symmetric around $y_\mt{plasma}$.  Interestingly, particle production in d+A collisions at RHIC \cite{Back:2003hx} seems consistent with both of these features, as already noted in \cite{Steinberg:2007fg}. 

At weak coupling we may determine $y_\mt{max}$ via perturbative QCD. For collisions with a large rapidity gap, $\left|y_\mt{A}-y_\mt{B}\right|\gg1$, this can be estimated  by equating the squared saturation scales of both colliding objects \cite{Xiao:2005pc}, $Q^2_s(N_\mt{A},y_\mt{max})=Q^2_s(N_\mt{B},y_\mt{max})$. Far from its own rapidity $y_\mt{C}$,    
the  saturation scale of a nucleus with $N_\mt{C}$ participating nucleons evolves as $Q^2_s(N_\mt{C},y) \sim N_\mt{C} 
\exp\left( \bar{\lambda} \left| y-y_\mt{C} \right| \right)$ \cite{Kharzeev:2001gp,Kharzeev:2004if}. 
The coupling-dependent exponent $\bar \lambda$ can be extracted  from fits to HERA data   within the saturation framework  \cite{Stasto:2000er} and is given by $\bar \lambda \simeq 0.25$ \cite{Kharzeev:2001gp,Kharzeev:2004if}, reflecting the fact that in perturbative QCD the fraction of energy available for particle production decreases with energy.  Substituting in the equation for $y_\mt{max}$ we  find \mbox{$y_\mt{max}=\frac{1}{2 \bar \lambda} \,\log(N_\mt{A}/N_\mt{B}) + y_\mt{NN}$}. We expect this estimate to be a better approximation for the LHC than for RHIC because of the much larger rapidity window of the former \cite{detailed}. 

We thus conclude that the strong- and weak-coupling predictions for $y_\mt{max}$ in p+A collisions differ by about a factor of 4 \cite{similar}. This makes the possible experimental extraction of  $y_\mt{max}$ from LHC p+A  data \cite{ALICE:2012xs} extremely interesting, since the result may help constrain the mechanism of bulk-matter production.

\noindent
{{\bf Acknowledgements.}} We thank P.~Steinberg for a suggestion that motivated this work,  A.~H.~Mueller for his explanation of particle production in perturbative QCD, and A.~Dumitru and P.~Bozek for useful discussions.
JCS and DM acknowledge financial support from grants FPA2010-20807 and CPAN CSD 2007-00042 Consolider-Ingenio 2010. JCS is further supported by a RyC fellowship and by grants 2009SGR502 and 
FP7-PEOPLE-2012-GIG-333786. DM is also supported by grants ERC StG HoloLHC-306605 and 2009SGR168. 
MPH is supported by the Netherlands Organization for Scientific
Research under the NWO Veni scheme (UvA) and by the National Science
Centre under Grant No. 2012/07/B/ST2/03794 (NCNR).
 WS is supported by a Utrecht University Foundations of Science grant and a Visiting Graduate Fellowship of Perimeter Institute.



\begin{thebibliography}{99}

\bibitem{Adcox:2004mh} 
  K.~Adcox {\it et al.}  [PHENIX Collaboration],
  Nucl.\ Phys.\ A {\bf 757}, 184 (2005)
  [nucl-ex/0410003].

\bibitem{Adams:2005dq} 
  J.~Adams {\it et al.}  [STAR Collaboration],
  Nucl.\ Phys.\ A {\bf 757}, 102 (2005)
  [nucl-ex/0501009].

\bibitem{Muller:2012zq} 
  B.~Muller, J.~Schukraft and B.~Wyslouch,
  Ann.\ Rev.\ Nucl.\ Part.\ Sci.\  {\bf 62}, 361 (2012)
  [arXiv:1202.3233 [hep-ex]].


\bibitem{Chesler:2010bi} 
  P.~M.~Chesler and L.~G.~Yaffe,
  Phys.\ Rev.\ Lett.\  {\bf 106}, 021601 (2011)
  [arXiv:1011.3562 [hep-th]].
  
\bibitem{Casalderrey-Solana:2013aba}
  J.~Casalderrey-Solana, M.~P.~Heller, D.~Mateos and W.~van der Schee,
  Phys.\ Rev.\ Lett.\  {\bf 111}, 181601 (2013)
  [arXiv:1305.4919 [hep-th]].
  
\bibitem{vanderSchee:2014qwa}
  W.~van der Schee,
 PhD thesis,  
  arXiv:1407.1849 [hep-th].

\bibitem{Chesler:2013lia} 
  P.~M.~Chesler and L.~G.~Yaffe,
  arXiv:1309.1439 [hep-th].



\bibitem{CasalderreySolana:2011us} 
  J.~Casalderrey-Solana, H.~Liu, D.~Mateos, K.~Rajagopal and U.~A.~Wiedemann,
  arXiv:1101.0618 [hep-th].

  
\bibitem{Aad:2013fja} 
  G.~Aad {\it et al.}  [ATLAS Collaboration],
  Phys.\ Lett.\ B {\bf 725}, 60 (2013)
  [arXiv:1303.2084 [hep-ex]].

\bibitem{Chatrchyan:2013nka} 
  S.~Chatrchyan {\it et al.}  [CMS Collaboration],
  Phys.\ Lett.\ B {\bf 724}, 213 (2013)
  [arXiv:1305.0609 [nucl-ex]].
  
  
\bibitem{Abelev:2012ola} 
  B.~Abelev {\it et al.}  [ALICE Collaboration],
  Phys.\ Lett.\ B {\bf 719}, 29 (2013)
  [arXiv:1212.2001].


\bibitem{Adare:2013piz} 
  A.~Adare {\it et al.}  [PHENIX Collaboration],
  [arXiv:1303.1794 [nucl-ex]].
  
\bibitem{Sickles:2013yna} 
  A.~M.~Sickles,
  arXiv:1309.6924 [nucl-th].



    
\bibitem{Bozek:2013uha} 
  P.~Bozek and W.~Broniowski,
  Phys.\ Rev.\ C {\bf 88}, 014903 (2013)
  [arXiv:1304.3044 [nucl-th]].
  
 \bibitem{footnote1} See \cite{Albacete:2009ji} for  previous related work.

  
\bibitem{Albacete:2009ji} 
  J.~L.~Albacete, Y.~V.~Kovchegov and A.~Taliotis,
  JHEP {\bf 0905}, 060 (2009)
  [arXiv:0902.3046 [hep-th]].
 
\bibitem{footnote2} We employ as a regulator a background thermal bath
with between 50 (7) and 20 (5) times less energy than the peak shock energy density for (a-)symmetric shock collisions. Each collision is simulated with three  different regulators, which allows us to check convergence and extrapolate to a collision without regulator.

\bibitem{Figueras:2009iu} 
  P.~Figueras, V.~E.~Hubeny, M.~Rangamani and S.~F.~Ross,
  JHEP {\bf 0904}, 137 (2009)
  [arXiv:0902.4696 [hep-th]].

\bibitem{Bzdak:2012tp}
  A.~Bzdak and D.~Teaney,
  Phys.\ Rev.\ C {\bf 87},  024906 (2013)
  [arXiv:1210.1965 [nucl-th]].


\bibitem{Back:2002wb}
  B.~B.~Back, M.~D.~Baker, D.~S.~Barton, R.~R.~Betts, M.~Ballintijn, A.~A.~Bickley, R.~Bindel and A.~Budzanowski {\it et al.},
  Phys.\ Rev.\ Lett.\  {\bf 91} (2003) 052303
  [nucl-ex/0210015].

\bibitem{Back:2004mr}
  B.~B.~Back {\it et al.}  [PHOBOS Collaboration],
  Phys.\ Rev.\ C {\bf 72} (2005) 031901
  [nucl-ex/0409021].


\bibitem{Back:2003hx} 
  B.~B.~Back {\it et al.}  [PHOBOS Collaboration],
  Phys.\ Rev.\ Lett.\  {\bf 93}, 082301 (2004)
  [nucl-ex/0311009].
  
\bibitem{Steinberg:2007fg} 
  P.~Steinberg,
  [nucl-ex/0703002 [NUCL-EX]].


  
\bibitem{Xiao:2005pc} 
  B.~-W.~Xiao,
  Phys.\ Rev.\ C {\bf 72}, 034905 (2005)
  [hep-ph/0505003].








\bibitem{Kharzeev:2001gp} 
  D.~Kharzeev and E.~Levin,
  Phys.\ Lett.\ B {\bf 523}, 79 (2001)
  [nucl-th/0108006].
  
\bibitem{Kharzeev:2004if} 
  D.~Kharzeev, E.~Levin and M.~Nardi,
  Nucl.\ Phys.\ A {\bf 747}, 609 (2005)
  [hep-ph/0408050].
  
\bibitem{Stasto:2000er} 
  A.~M.~Stasto, K.~J.~Golec-Biernat and J.~Kwiecinski,
  Phys.\ Rev.\ Lett.\  {\bf 86}, 596 (2001)
  [hep-ph/0007192].
  
\bibitem{detailed}
A more precise estimate may be obtained from a detailed analysis \cite{Dumitru:2011wq,Albacete:2012xq} that includes the effects of \mbox{large-$x$} physics, but this is not necessary for our purposes. 
   
\bibitem{similar}  
An analogous but smaller difference arises in A+A collisions due to event-by-event fluctuations in the number of participating nucleons.

\bibitem{ALICE:2012xs} 
  B.~Abelev {\it et al.}  [ALICE Collaboration],
  Phys.\ Rev.\ Lett.\  {\bf 110}, 032301 (2013)
  [arXiv:1210.3615 [nucl-ex]].
  
\bibitem{Dumitru:2011wq} 
  A.~Dumitru, D.~E.~Kharzeev, E.~M.~Levin and Y.~Nara,
  Phys.\ Rev.\ C {\bf 85}, 044920 (2012)
  [arXiv:1111.3031 [hep-ph]].
  
\bibitem{Albacete:2012xq} 
  J.~L.~Albacete, A.~Dumitru, H.~Fujii and Y.~Nara,
  Nucl.\ Phys.\ A {\bf 897}, 1 (2013)
  [arXiv:1209.2001 [hep-ph]].
  


  


\end{thebibliography}

\end{document}